\documentclass[prb,superscriptaddress,showpacs,floatfix,tightenlines,10pt,twocolumn]{revtex4}
\usepackage{amssymb}
\usepackage{amsmath}
\usepackage{graphicx}

\begin{document}

\title{Zeeman coupling and screening corrections to skyrmion excitations in
graphene}
\author{Wenchen Luo}
\affiliation{D\'{e}partement de physique, Universit\'{e} de Sherbrooke, Sherbrooke, Qu%
\'{e}bec, J1K 2R1, Canada}
\author{R. C\^{o}t\'{e} }
\affiliation{D\'{e}partement de physique, Universit\'{e} de Sherbrooke, Sherbrooke, Qu%
\'{e}bec, J1K 2R1, Canada}
\keywords{skyrmion,graphene,transport gap}
\pacs{73.50.Fq,72.10.-d,73.21.-b}

\begin{abstract}
At half filling of the fourfold degenerate Landau levels $\left\vert
n\right\vert \geq 1$ in graphene, the ground states are spin polarized
quantum Hall states that support spin skyrmion excitations for $\left\vert
n\right\vert =1,2,3$. Working in the Hartree-Fock approximation, we compute
the excitation energy of an unbound spin skyrmion-antiskyrmion excitation as
a function of the Zeeman coupling strength for these Landau levels. We find
for both the bare and screened Coulomb interactions that the spin
skyrmion-antiskyrmion excitation energy is lower than the excitation energy
of an unbound spin $1/2$ electron-hole pair in a finite range of Zeeman
coupling in Landau levels $\left\vert n\right\vert =1,2,3$. This range
decreases rapidly for increasing Landau level index and is extremely small
for $\left\vert n\right\vert =3.$ For valley skyrmions which should be
present at $1/4$ and $3/4$ fillings of the Landau levels $\left\vert
n\right\vert =1,2,3$, we show that screening corrections are more important
in the latter case. It follows that an unbound valley skyrmion-antiskyrmion
excitation has lower energy at $3/4$ filling than at $1/4.$ We compare our
results with recent experiments on spin and valley skyrmion excitations in
graphene.
\end{abstract}

\date{\today }
\maketitle

\section{ INTRODUCTION}

The energy of the Landau levels in graphene in a transverse magnetic field $%
\mathbf{B}=-B\widehat{\mathbf{z}}$ is given by%
\begin{equation}
E_{s,n}^{0}=sgn\left( n\right) \sqrt{\left\vert n\right\vert }\frac{\sqrt{2}%
\hslash v_{F}}{\ell }-\frac{1}{2}sg\mu _{B}B,
\end{equation}%
where $n=0,\pm 1,\pm 2,...$ is the Landau level index, $s=\pm 1$ is the spin
index, $\ell =\sqrt{\hslash c/eB}$ is the magnetic length, $v_{F}$ is the
Fermi velocity, $g=2$ is the Land\'{e} factor and $\mu _{B}$ is the Bohr
magneton. Because the Zeeman energy is very small in comparison with the
kinetic energy, each Landau level is usually considered as being fourfold
degenerate when counting spin and valley ($K_{\pm }$) degrees of freedom.
Experimentally, this kinetic energy quantization leads to the anomalous Hall
sequence\cite{Graphenereview}%
\begin{equation}
\sigma _{xy}=\pm \frac{4e^{2}}{h}\left( n+\frac{1}{2}\right)
\end{equation}%
in the Hall conductivity and so to quantum Hall plateaus at filling factors $%
\nu =\pm 4$ $\left( n+\frac{1}{2}\right) .$

In experiments\cite{Afyoung} on very high quality graphene samples
fabricated on hexagonal boron nitride (hBN) substrates, it is possible to
resolve the quantum Hall plateaus at \textit{all} integer filling factors,
i.e. $\sigma _{xy}=me^{2}/h$ with $\left\vert m\right\vert \geq 0$\cite{Kim}$%
,$ and to see an insulating state developing at filling factor $\nu =0.$
These experiments allow the study of the nature of the quantum Hall ground
states of the chiral two-dimensional electron gas (C2DEG) as well as the
nature of their charged excitations. In Ref. \onlinecite{Afyoung}, it was
shown that the ground states are maximally spin polarized at filling factors 
$\nu =-4,-8,-12$ while the ground state at $\nu =0$ is not. Moreover, the
charged excitations were found to be spin texture excitations at half
filling of Landau levels $n=-1,-2$ in some range of Zeeman coupling. Valley
skyrmions were also detected and studied at filling factors $\nu =-3,-5$
where the ground state is valley polarized.

Theoretically, a calculation based on the nonlinear $\sigma $ model ($%
NL\sigma $ model) that is valid at zero Zeeman coupling shows that, in
graphene, the transport gap should be due to spin texture i.e. spin skyrmion
excitations\cite{Sondhi,Note2} at half filling of Landau levels $\left\vert
n\right\vert =1,2,3$\cite{Kunyang} and to valley skyrmions at $1/4$ and $3/4$
filling of these same levels. Because there are no symmetry-breaking terms
associated with the two valleys (i.e. no equivalent Zeeman coupling), the $%
NL\sigma $ model calculation of Ref. \onlinecite{Kunyang} describes the
valley skyrmions very well. In graphene, spin skyrmions thus persist to
higher Landau levels than in a conventional semiconductor two-dimensional
electron gas (2DEG). Indeed, in a semiconductor 2DEG, skyrmions are the
lowest-energy charged excitations in $n=0$ at filling factor $\nu =1$ only
(when the width of the quantum well is neglected)\cite{Fertig,Fertig2}. In
higher Landau levels, the transport gap is due to unbound electron-hole pairs%
\cite{Wu}. The same conclusion concerning skyrmions in graphene was reached
using the density matrix renormalization group (DMRG) method for $n=0,1,2.$
For $n=3,$ the skyrmion-antiskyrmion (S-aS) pair and electron-hole pair
energies are very close and it was not possible to stabilize a skyrmion
solution with the DMRG method\cite{Shibata}. Exact diagonalization studies
of valley skyrmions have also been done in Ref. \onlinecite{Toke2}. Crystals
of valley skyrmions have been shown to be the ground state of the C2DEG
around quarter filling of the $n=0,1$ Landau levels\cite{Jobidon}. A
theoretical study of the possible entanglement between the spin and valley
degrees of freedom in graphene which could lead to CP$^{3}$ skyrmions was
done in Ref. \onlinecite{Doucot}. This work did not include a calculation of
the behavior of the transport gap with Zeeman coupling however.

The energy functional of the $NL\sigma $ model in broken-symmetry quantum
Hall ferromagnetic states contains a gradient term that originates from the
exchange part of the Coulomb interaction. The topological solitons of this
model can be determined exactly\cite{Rajaraman}. The gradient term being
scale invariant in two dimensions, the energy of these solutions is
independent of their size. When a finite Zeeman coupling is considered, two
more terms must be added to the $NL\sigma $ model energy functional: a
Zeeman coupling which favors small skyrmions and an electrostatic
self-interaction energy which favors large skyrmions. These two terms
compete together to determine the optimal size and energy of a skyrmion as
well as its density profile and spin texture.

In this paper, we study the energetics of spin skyrmions in graphene in the
half-filled Landau levels $\left\vert n\right\vert =1,2,3$ and
valley-skyrmions at $1/4$ and $3/4$ fillings where the ground state is spin
and valley polarized. For spin skyrmions, we extend the calculation of the $%
NL\sigma $ model, valid at zero Zeeman coupling, to finite Zeeman coupling
by using a Green's function approach. The equation of motion of the Green's
function is derived in the Hartree-Fock approximation and in the symmetric
gauge. This derivation leads to a set of coupled self-consistent equations
for the angular momentum components of the skyrmion wave function which must
be solved numerically using an iterative procedure. Our approach is
equivalent to the canonical transformation method used earlier in the study
of skyrmion in a semiconductor 2DEG\cite{Fertigskyrmion}. Our method works
well with finite-size skyrmions but cannot deal with very large skyrmions
which are obtained at small Zeeman coupling since large skyrmions require a
large number of angular momentum components for their description.

We compute the unbound S-aS pair energy, $\Delta _{S-aS},$ and compare it
with the energy to make an unbound electron-hole pair $\Delta _{e-h}.$ The
transport gap is determined by the lowest of these two energies. We find
that spin-texture excitations are the lowest-energy excitations at
half-filling in a small range of Zeeman coupling for $n=1,2$ and that this
range decreases rapidly with increasing Landau level index. According to the 
$NL\sigma $ model, spin skyrmions are the lowest-energy excitations at zero
Zeeman coupling also in $n=3.$ With the limitations of our method, however,
we cannot find spin skyrmions at finite Zeeman coupling for $n=3$. It
follows that the Zeeman coupling range where they are the lowest-energy
excitations must be very small.

It is straightforward to modify our method to include screening of the
Coulomb interaction. The Coulomb matrix elements that enter the equation of
motion for the Green's functions are evaluated using a dielectric function
computed in the random-phase approximation (RPA). For spin skyrmions, we
find that screening decreases substantially the transport gaps $\Delta
_{NL\sigma M},\Delta _{S-aS}$ and $\Delta _{e-h}$ as well as the critical
Zeeman coupling for the transition between $\Delta _{S-aS}$ and $\Delta
_{e-h}.$ Nevertheless, the skyrmion scenario still prevails for $n=1,2$ and
for $n=3$ in a very small range of Zeeman coupling.

For valley skyrmions, there is no symmetry-breaking term equivalent to the
Zeeman coupling so that the transport gap can be computed using the $%
NL\sigma M.$ Our results show that screening corrections are more important
at $3/4$ filling than at $1/4$ so that the transport gap due to unbound
valley S-aS excitations is lower in the former case.

This paper is organized in the following way. In Sec. II, we introduce the
spinor non-interacting electronic states of graphene in a magnetic field
using the symmetric gauge. Sec. III summarizes the Hartree-Fock
approximation to the electron-electron interaction and presents the
assumptions necessary to derive the two-level system that is the starting
point of our work. Sec. IV contains a description of the quasiparticle
(electron and hole) and skyrmion excitations. The Green's function formalism
is discussed in Sec. V. The numerical results are presented and discussed in
Sec. VI for the unscreened Coulomb interaction and in Sec. VII for the
screened interaction. We conclude in Sec. VIII.

\section{TIGHT-BINDING HAMILTONIAN AND EIGENSTATES OF THE NON-INTERACTING
CHIRAL 2DEG}

Graphene has a honeycomb lattice structure that can be described as an
hexagonal Bravais lattice with a lattice constant $a_{0}=2.46$ \AA\ and a
basis of two carbon atoms $A$ and $B$\cite{Graphenereview}. Each carbon atom
contributes one electron to the two $\pi $ bands. These electrons form a
C2DEG. In the sublattice basis $\left( A,B\right) $, the Hamiltonian in a
transverse magnetic field $\mathbf{B}=-B\widehat{\mathbf{z}}$ and in the
continuum approximation (i.e. for small energy with respect to the Dirac
points) is given by%
\begin{equation}
\mathcal{H}_{\alpha }=\alpha \frac{\sqrt{2}\hslash v_{F}}{\ell }\left( 
\begin{array}{cc}
0 & a^{\mp } \\ 
a^{\pm } & 0%
\end{array}%
\right) ,  \label{w20}
\end{equation}%
where $\alpha =\pm $ is the valley index for the two nonequivalent valleys $%
\mathbf{K}_{\alpha }=\alpha \left( 2/3,0\right) \left( 2\pi /a_{0}\right) $
in the Brillouin zone and $v_{F}=\sqrt{3}\gamma _{0}a_{0}/2\hslash $ is the
Fermi velocity with $\gamma _{0}=3.12$ eV the hopping energy between
nearest-neighbors carbon atoms. The operators $a^{+}=a^{\dag }$,$a^{-}=a$
are the ladder operators for the one-dimensional harmonic oscillator. In Eq.
(\ref{w20}), the upper(lower) sign is for the $\alpha =+\left( -\right) $
valley.

The Landau level spectrum of $\mathcal{H}_{\alpha }$ is given by%
\begin{equation}
E_{s,n}^{0}=sgn\left( n\right) \sqrt{\left\vert n\right\vert }\frac{\sqrt{2}%
\hslash v_{F}}{\ell }-\frac{1}{2}s\Delta _{Z}.  \label{w21}
\end{equation}%
In Eq. (\ref{w21}), we have added a Zeeman coupling $\Delta _{Z}=g\mu _{B}B$
to $\mathcal{H}_{\alpha }$. The Landau level index $n=0,\pm 1,\pm 2,...$
takes both positive and negative values. In the absence of Zeeman coupling,
each Landau level is fourfold degenerate when counting valley and spin
degrees of freedom. In addition, each Landau level has the macroscopic
orbital degeneracy $N_{\varphi }=S/2\pi \ell ^{2}$ where $S$ is the C2DEG
area.

The eigenstates of $\mathcal{H}_{\alpha }$ are spinors in the sublattice
basis $\left( A,B\right) .$ For $n\neq 0,$ these spinors are for the two
valleys and for a given spin orientation, given by 
\begin{eqnarray}
\left\vert n,m,\alpha =+\right\rangle &=&\frac{1}{\sqrt{2}}\left( 
\begin{array}{c}
sgn\left( n\right) \left\vert \left\vert n\right\vert -1,m\right\rangle \\ 
\left\vert \left\vert n\right\vert ,m\right\rangle%
\end{array}%
\right) ,  \label{w22} \\
\left\vert n,m,\alpha =-\right\rangle &=&\frac{1}{\sqrt{2}}\left( 
\begin{array}{c}
\left\vert \left\vert n\right\vert ,m\right\rangle \\ 
-sgn\left( n\right) \left\vert \left\vert n\right\vert -1,m\right\rangle%
\end{array}%
\right) ,  \label{w23}
\end{eqnarray}%
while for $n=0,$ the eigenspinors are given by 
\begin{eqnarray}
\left\vert 0,m,\alpha =+\right\rangle &=&\left( 
\begin{array}{c}
0 \\ 
\left\vert 0,m\right\rangle%
\end{array}%
\right) ,  \label{w24} \\
\left\vert 0,m,\alpha =-\right\rangle &=&\left( 
\begin{array}{c}
\left\vert 0,m\right\rangle \\ 
0%
\end{array}%
\right) .  \label{w25}
\end{eqnarray}%
There is a direct correspondence between valley and sublattice indices in
Landau level $n=0.$

The spinors in Eqs. (\ref{w22}-\ref{w25}) are written in the symmetric gauge 
$\mathbf{A}=\left( By/2,-Bx/2\right) ,$ where the quantum number $%
m=0,1,2,3,...$ is associated with the angular momentum by the relation%
\begin{equation}
L_{z}\left\vert n,m\right\rangle =(m-n)\hslash \left\vert n,m\right\rangle .
\end{equation}%
The states $\left\vert n,m\right\rangle $ are simply the eigenstates of a
conventional (non chiral) 2DEG in a magnetic field. Because of the symmetry
of the skyrmion charged excitations, the symmetric gauge is the most
convenient one.

In real space, the corresponding wave functions are given by\cite{Yoshioka}%
\begin{eqnarray}
\varphi _{n,m}\left( \mathbf{r}\right) &\equiv &\left\langle \mathbf{r}%
|n,m\right\rangle \\
&=&B_{n,m}e^{i\left( m-n\right) \phi }\left( \frac{r}{\ell }\right)
^{\left\vert m-n\right\vert }e^{-\frac{r^{2}}{4\ell ^{2}}}  \notag \\
&&\times L_{\frac{n+m}{2}-\frac{\left\vert n-m\right\vert }{2}}^{\left\vert
m-n\right\vert }\left( \frac{r^{2}}{2\ell ^{2}}\right) ,  \notag
\end{eqnarray}%
where $\phi $ is the angle between the vector $\mathbf{r}$ and the $x$ axis, 
$L_{n}^{m}\left( x\right) $ is a generalized Laguerre polynomial and the
normalization constant is given by%
\begin{equation}
B_{n,m}=\frac{C_{n,m}\left( -i\right) ^{n}}{\sqrt{2^{\left\vert
m-n\right\vert +1}\pi \ell ^{2}}}\sqrt{\frac{\left( \frac{n+m}{2}-\frac{%
\left\vert n-m\right\vert }{2}\right) !}{\left( \frac{n+m}{2}+\frac{%
\left\vert n-m\right\vert }{2}\right) !}}
\end{equation}%
with $C_{n,m}=1$ for $m\leq n$ and $C_{n,m}=\left( -1\right) ^{m-n}$ for $%
m>n.$

\section{HARTREE-FOCK APPROXIMATION TO THE INTERACTING CHIRAL\ 2DEG}

In this paper, we consider the situation where the quartet of states in
Landau level $n$ is partially filled and levels $n^{\prime }<n$ are
completely filled. We make the approximation of considering the filled
levels as inert so that we can ignore them altogether. We thus neglect
Landau level mixing. It must be kept in mind, however, that in graphene the
difference in the kinetic energy between the first two Landau levels$~E_{C}=%
\sqrt{2}\hslash v_{F}/\ell =3.\,\allowbreak 67\times 10^{-2}\sqrt{B}$eV$=426%
\sqrt{B}$ K is of the order of the Coulomb interaction $e^{2}/\kappa \ell
=2.25\times 10^{-2}\sqrt{B}$ eV$=261\sqrt{B}$ K (for $\kappa =2.5$
appropriate for graphene on hexagonal boron nitride and $B$ in Tesla) and
Landau level mixing may be important. The Zeeman energy $\Delta _{Z}=g\mu
_{B}B=1.\,\allowbreak 16\times 10^{-4}B$ eV$=1.34B$ K.

We need to consider the Coulomb interaction between electrons in level $n$
which is given in second-quantization by

\begin{eqnarray}
V &=&\frac{1}{2}\underset{\alpha ,\beta ,s,s^{\prime }}{\sum }\int d\mathbf{r%
}\int d\mathbf{r}^{\prime }\Psi _{n,s,\alpha }^{\dagger }(\mathbf{r})\Psi
_{n,s^{\prime },\beta }^{\dagger }(\mathbf{r}^{\prime })  \label{w1} \\
&&\times V(\mathbf{r}-\mathbf{r}^{\prime })\Psi _{n,s^{\prime },\beta }(%
\mathbf{r}^{\prime })\Psi _{n,s,\alpha }(\mathbf{r}),  \notag
\end{eqnarray}%
where the Coulomb potential $V\left( \mathbf{r}\right) =e^{2}/\kappa r$ with 
$\kappa $ the dielectric constant of the substrate holding the graphene
layer. The electron (spinor-)field operator is written as 
\begin{equation}
\Psi _{n,s,\alpha }(\mathbf{r})=\sum_{m}\left\langle \mathbf{r}|n,m,\alpha
\right\rangle c_{s,\alpha ,n,m},
\end{equation}%
where $c_{s,\alpha ,n,m}$ annihilates an electron of spin $s$ in valley $%
\alpha ,$ Landau level $n,$ and orbital quantum number $m.$ In Eq. (\ref{w1}%
), the terms that do not conserve the valley index are very small and have
been neglected\cite{Goerbig1}.

To ensure the system's neutrality, an interaction between the C2DEG and a
uniform positive background of density $n_{b}=N_{e}/S$ where $N_{e}$ is the
number of electrons in level $n$ must be added to $V$. That interaction is
given by 
\begin{equation}
V_{e-b}=-n_{b}N_{e}\int d\mathbf{r}V\left( \mathbf{r}\right) ,
\end{equation}%
where 
\begin{equation}
N_{e}=\sum_{\alpha ,s}\int d\mathbf{r}\Psi _{n,s,\alpha }^{\dag }\left( 
\mathbf{r}\right) \Psi _{n,s,\alpha }\left( \mathbf{r}\right) .
\end{equation}

Making the usual Hartree-Fock pairing of the field operators in Eq. (\ref{w1}%
), we get for the Hartree-Fock Hamiltonian%
\begin{eqnarray}
H_{HF} &=&\sum_{s,\alpha ,m}E_{s,n}^{0}c_{s,\alpha ,n,m}^{\dag }c_{s,\alpha
,n,m}  \label{w3} \\
&&+\sum_{s,s^{\prime }}\sum_{\alpha ,\beta }\sum_{m_{1}\ldots
m_{4}}V_{m_{1},m_{2},m_{3},m_{4}}^{n}  \notag \\
&&\times \left\langle c_{s,\alpha ,n,m_{1}}^{\dag }c_{s,\alpha
,n,m_{2}}\right\rangle c_{s^{\prime },\beta ,n,m_{3}}^{\dag }c_{s^{\prime
},\beta ,n,m_{4}}  \notag \\
&&-\sum_{s,s^{\prime }}\sum_{\alpha ,\beta }\sum_{m_{1}\ldots
m_{4}}V_{m_{1},m_{2},m_{3},m_{4}}^{n}  \notag \\
&&\times \left\langle c_{s,\alpha ,n,m_{1}}^{\dag }c_{s^{\prime },\beta
,n,m_{4}}\right\rangle c_{s^{\prime },\beta ,n,m_{3}}^{\dag }c_{s,\alpha
,n,m_{2}}  \notag \\
&&-\nu _{n}\sum_{s,\alpha
,m_{1},m_{2}}V_{m_{1},m_{1},m_{2},m_{2}}^{n}c_{s,\alpha ,n,m_{2}}^{\dag
}c_{s,\alpha ,n,m_{2}},  \notag
\end{eqnarray}%
where the last term is the interaction with the positive background.

The interactions $V_{m_{1},m_{2},m_{3},m_{4}}^{n}$ in Eq. (\ref{w3}) are
defined by 
\begin{eqnarray}
&&V_{m_{1},m_{2},m_{3},m_{4}}^{n}  \label{w2} \\
&=&V_{m_{1},m_{2},m_{3},m_{4}}^{0,0,0,0}\delta _{n,0}  \notag \\
&&+\frac{1}{4}\left[ V_{m_{1},m_{2},m_{3},m_{4}}^{\left\vert n\right\vert
,\left\vert n\right\vert ,\left\vert n\right\vert ,\left\vert n\right\vert
}+V_{m_{1},m_{2},m_{3},m_{4}}^{\left\vert n\right\vert -1,\left\vert
n\right\vert -1,\left\vert n\right\vert -1,\left\vert n\right\vert -1}\right]
\Theta \left( \left\vert n\right\vert \right)  \notag \\
&&+\frac{1}{4}\left[ V_{m_{1},m_{2},m_{3},m_{4}}^{\left\vert n\right\vert
,\left\vert n\right\vert ,\left\vert n\right\vert -1,\left\vert n\right\vert
-1}+V_{m_{1},m_{2},m_{3},m_{4}}^{\left\vert n\right\vert -1,\left\vert
n\right\vert -1,\left\vert n\right\vert ,\left\vert n\right\vert }\right]
\Theta \left( \left\vert n\right\vert \right) ,  \notag
\end{eqnarray}%
where%
\begin{eqnarray}
V_{m_{1},m_{2},m_{3},m_{4}}^{n_{1},n_{2},n_{3},n_{4}} &=&\int d\mathbf{r}%
\varphi _{n_{1},m_{1}}^{\ast }\left( \mathbf{r}\right) \varphi
_{n_{2},m_{2}}\left( \mathbf{r}\right)  \label{elements} \\
&&\times \int d\mathbf{r}^{\prime }\frac{e^{2}}{\kappa \left\vert \mathbf{r}-%
\mathbf{r}^{\prime }\right\vert }\varphi _{n_{3},m_{3}}^{\ast }\left( 
\mathbf{r}^{\prime }\right) \varphi _{n_{4},m_{4}}\left( \mathbf{r}^{\prime
}\right) .  \notag
\end{eqnarray}%
The matrix elements that are needed in Eq. (\ref{w2}) are all of the form $%
V_{m_{1},m_{2},m_{3},m_{4}}^{n,n,q,q}$ and can be evaluated numerically
using the following expression%
\begin{eqnarray}
&&V_{m_{1},m_{2},m_{3},m_{4}}^{n,n,q,q}  \label{vm1m2} \\
&=&\left( \frac{e^{2}}{\kappa \ell }\right) \sqrt{\frac{Min\left(
m_{1},m_{2}\right) !}{Max\left( m_{1},m_{2}\right) !}}\sqrt{\frac{Min\left(
m_{3},m_{4}\right) !}{Max\left( m_{3},m_{4}\right) !}}  \notag \\
&&\times \delta _{m_{1}+m_{3},m_{2}+m_{4}}\sqrt{2}\int_{0}^{\infty
}dxe^{-2x^{2}}x^{2\left\vert m_{1}-m_{2}\right\vert }  \notag \\
&&\times L_{n}^{0}\left( x^{2}\right) L_{q}^{0}\left( x^{2}\right)
L_{Min\left( m_{1},m_{2}\right) }^{\left\vert m_{1}-m_{2}\right\vert }\left(
x^{2}\right) L_{Min\left( m_{3},m_{4}\right) }^{\left\vert
m_{3}-m_{4}\right\vert }\left( x^{2}\right) .  \notag
\end{eqnarray}

The Hamiltonian of Eq. (\ref{w3}) is very general and allows the calculation
of skyrmion excitations with valley pseudospin texture, spin texture or even
skyrmions with intertwined spin and valley pseudospin textures. In this
paper, we restrict ourselves to situations where the quartet of state in
Landau level $\left\vert n\right\vert >0$ is half-filled in which case the
ground is spin polarized and spin-skyrmions excitations are possible and to $%
1/4$ or $3/4$ fillings in which cases the ground state is valley polarized
and valley skyrmions are possible. Experiments show that the ground states
in $n=0$ are more complex\cite{Afyoung} and we will not consider this Landau
level. Indeed, for $\nu =0,$ the ground state is probably not fully spin
polarized and the nature of the broken-symmetry ground state is still debated%
\cite{Debate}. At $\nu =-1,$ experiments suggest that excitations contain
both valley and spin flips.

At half filling, the states with up spins in both valleys are occupied. In a
spin skyrmion excitation, an electron of spin $s=-1$ is added to the ground
state and causes a certain number of spins $s=+1$ to flip to the $s=-1$
state in order to minimize the Coulomb exchange energy between electrons.
These spins reversal can, in principle, occur in both valleys. But, because
of the SU(2) valley symmetry of the Hamiltonian of Eq. (\ref{w3}), it is
equivalent to consider that they originate from one of the valley only. When
we do so, we assume that the other valley plays no role and can be
considered as inert. When considering spin skyrmions only, we can thus
restrict the Hilbert space in Landau level $n$ to one valley, say $\alpha
=+1,$ and to two spin orientations. In this way, we can work with a two-
instead of a four-level system. The same principle can be applied to the
ground state at $1/4$ or $3/4$ fillings. At $1/4$ filling, for example,
state of up spins in valley $K_{+}$ (or any linear combination of $K_{+}$
and $K_{-}$because of the SU(2) valley symmetry) are occupied. If we assume
that spin flips are not possible because of the finite Zeeman coupling, than
we have a two-level system with $K_{\pm }$ and up spins and excitations are
valley skyrmions. At $3/4$ filling, we have a two-level system with states $%
K_{\pm }$ and down spins.

From now on, we drop the Landau level indices (except in the interaction $%
V^{n}$) and write the Hartree-Fock Hamiltonian as

\begin{eqnarray}
H_{HF} &=&\sum_{s,m}E_{s}^{0}\rho _{m,m}^{s,s}  \label{w10} \\
&&+\frac{1}{2}\sum_{s,s^{\prime }}\sum_{m_{1}\ldots
m_{4}}V_{m_{1},m_{2},m_{3},m_{4}}^{n}\left\langle \rho
_{m_{1},m_{2}}^{s,s}\right\rangle \rho _{m_{3},m_{4}}^{s^{\prime },s^{\prime
}}  \notag \\
&&-\frac{1}{2}\sum_{s,s^{\prime }}\sum_{m_{1}\ldots
m_{4}}V_{m_{1},m_{2},m_{3},m_{4}}^{n}\left\langle \rho
_{m_{1},m_{4}}^{s,s^{\prime }}\right\rangle \rho _{m_{3},m_{2}}^{s^{\prime
},s}  \notag \\
&&-\nu _{n}\sum_{s,m_{1},m_{2}}V_{m_{1},m_{1},m_{2},m_{2}}^{n}\left\langle
\rho _{m_{2},m_{2}}^{s,s}\right\rangle  \notag \\
&&+\frac{1}{2}\nu _{n}^{2}\sum_{m_{1},m_{2}}V_{m_{1},m_{1},m_{2},m_{2}}^{n},
\notag
\end{eqnarray}%
where we have defined the operator%
\begin{equation}
\rho _{m_{1},m_{2}}^{s,s^{\prime }}=c_{s,m_{1}}^{\dag }c_{s^{\prime },m_{2}}.
\end{equation}%
The last term in Eq. (\ref{w10}) is the background's electrostatic
interaction $V_{b-b}=\frac{1}{2}n_{b}^{2}\int d\mathbf{r}\int d\mathbf{r}%
^{\prime }V\left( \mathbf{r}-\mathbf{r}^{\prime }\right) $ which must be
included in $H_{HF}$ in order to correctly take into account the system's
neutrality when computing excitation energies.

The Hamiltonian $H_{HF}$ contains the $s$ index and is written with spin
skyrmions in mind. We give all the subsequent formulas for spin skyrmions.
Valley skyrmions are easily treated by replacing $s$ with the valley index $%
\alpha $ in these formulas and neglecting the Zeeman term in the excitations
energy.

We remark that, for $n=0,$ the interactions $V_{m_{1},m_{2},m_{3},m_{4}}^{n}$
given by Eq. (\ref{w2}) are identical to that of a conventional 2DEG. If one
assumes a spin-polarized ground state for $n=0$ the quasiparticle and
spin-skyrmion excitation energies found from Eq. (\ref{w10}) are identical
to those of a non-chiral 2DEG's which were computed in Refs. %
\onlinecite{Sondhi, Fertigskyrmion}.

\section{QUASIPARTICLE AND SKYRMION EXCITATIONS}

In the two-level system, the ground state is given by

\begin{equation}
\left\vert GS\right\rangle =\prod\limits_{m}^{\infty}c_{+,m}^{\dag
}\left\vert 0\right\rangle
\end{equation}%
which implies that 
\begin{equation}
\left\langle \rho _{m,m^{\prime }}^{s,s^{\prime }}\right\rangle =\delta
_{m,m^{\prime }}\delta _{s,s^{\prime }}\delta _{s,+}.
\end{equation}%
Its energy is given by 
\begin{equation}
E_{GS}=-\frac{1}{2}N_{e}g\mu _{B}B-\frac{1}{2}%
\sum_{m_{1},m_{2}}V_{m_{1},m_{2},m_{2},m_{1}}^{n}.  \label{gs}
\end{equation}%
\newline

An quasi-electron excitation is obtained by adding one electron of spin $%
s=-1 $ and angular momentum $m_{0}$ to the ground state i.e. 
\begin{equation}
\left\vert e\right\rangle =c_{-,m_{0}}^{\dag }\left\vert GS\right\rangle
\end{equation}%
and has%
\begin{equation}
\left\langle \rho _{m,m^{\prime }}^{s,s^{\prime }}\right\rangle =\left\{ 
\begin{array}{ccc}
\delta _{s,s^{\prime }}\delta _{m,m^{\prime }}, & \mathrm{if} & s=+1; \\ 
\delta _{m,m_{0}}\delta _{s,s^{\prime }}\delta _{m,m^{\prime }}, & \mathrm{if%
} & s=-1.%
\end{array}%
\right.
\end{equation}%
The energy required to add one electron to the ground state is 
\begin{equation}
\Delta _{e}=\frac{1}{2}\Delta _{Z}
\end{equation}%
and is independent of the value of $m_{0}.$

For the quasi-hole state,%
\begin{equation}
\left\vert h\right\rangle =c_{+,m_{0}}\left\vert GS\right\rangle
\end{equation}%
with 
\begin{equation}
\left\langle \rho _{m,m^{\prime }}^{s,s^{\prime }}\right\rangle =\left\{ 
\begin{array}{ccc}
\delta _{s,s^{\prime }}\delta _{m,m^{\prime }}\left( 1-\delta
_{m,m_{0}}\right) , & \mathrm{if} & s=+1; \\ 
0, & \mathrm{if} & s=-1%
\end{array}%
\right.
\end{equation}%
and the energy required to create this state is given by 
\begin{equation}
\Delta _{h}=\frac{1}{2}\Delta _{Z}+\sum_{m}V_{m,m_{0},m_{0},m}^{n}.
\end{equation}%
If follows that the energy required to create an Hartree-Fock electron-hole
pair with both particles infinitely separated in space is given by%
\begin{eqnarray}
\Delta _{e-h} &=&\Delta _{e}+\Delta _{h}  \label{deltaeh} \\
&=&\Delta _{Z}+\sum_{m}V_{m,m_{0},m_{0},m}^{n}.  \notag
\end{eqnarray}%
Numerically, the value of $\Delta _{eh}$ is independent of the choice of $%
m_{0}.$ In fact%
\begin{equation}
\sum_{m}V_{m,m_{0},m_{0},m}^{n}=\left( \frac{e^{2}}{\kappa \ell }\right)
\int_{0}^{\infty }\frac{dx}{2\pi }\left\vert \Lambda _{n}\left( x\right)
\right\vert ^{2},  \label{ement}
\end{equation}%
with%
\begin{eqnarray}
\Lambda _{n}\left( x\right) &=&\delta _{n,0}e^{-x^{2}/4}+\frac{1}{2}\Theta
\left( \left\vert n\right\vert \right) e^{-x^{2}/4}  \label{lamn} \\
&&\times \left[ L_{\left\vert n\right\vert }^{0}\left( \frac{x^{2}}{2}%
\right) +L_{\left\vert n\right\vert -1}^{0}\left( \frac{x^{2}}{2}\right) %
\right] .  \notag
\end{eqnarray}

Following Ref. \onlinecite{Fertigskyrmion}, the skyrmion state in Landau
level $n$ is written as%
\begin{equation}
\left\vert S\right\rangle =\prod\limits_{p=0}^{\infty }\left(
u_{p}c_{+,p}^{\dag }+v_{p}c_{-,p+1}^{\dag }\right) c_{-,0}^{\dag }\left\vert
0\right\rangle ,  \label{sk}
\end{equation}%
with the constraint 
\begin{equation}
\left\vert u_{p}\right\vert ^{2}+\left\vert v_{p}\right\vert ^{2}=1.
\end{equation}%
This state has energy $E_{S}.$ The quasiparticle state $\left\vert
e\right\rangle =c_{-,0}^{\dag }\left\vert GS\right\rangle $ corresponds to
the limit $u_{p}=1$ and $v_{p}=0$ for all $p^{\prime }s$ i.e. to a zero-size
skyrmion. The skyrmion excitation energy is given by%
\begin{equation}
\Delta _{S}=E_{S}-E_{GS}.  \label{dsk}
\end{equation}

In the skyrmion state, one electron of spin down and quantum state $p=0$ is
added to the C2DEG and, at the same time, the state with spin up and quantum
number $p$ is combined with a state with spin down and quantum number $p+1.$
The difference in angular momentum between the two states is $\Delta
l_{z}=+\hslash $ and such pairing produces a $2\pi $ counter-clockwise
rotation of the spins in real space as shown in Fig. \ref{fig5}(a) below. It
is easy to show that the state $\left\vert S\right\rangle $ describes a spin
texture with a unit topological charge. The variational freedom in the wave
function of this state allows deviations of the spin texture from that of
the pure \textit{NL}$\sigma $ model. Far from the origin, this state is
locally identical to the ferromagnetic ground state and all spins point in
the \textquotedblleft up\textquotedblright\ direction. Near the origin, the
projection of the total spin along the field direction becomes negative. The
total increase (decrease) in the electron charge near the origin compared to
the ferromagnetic ground state corresponds to one added electron(hole) for
the skyrmion(antiskyrmion)

The total number of reversed spins in the skyrmion state is given by%
\begin{equation}
K=\sum_{p}\left\vert v_{p}\right\vert ^{2}.  \label{flip}
\end{equation}

In a similar way, the antiskyrmion state is given by 
\begin{equation}
\left\vert aS\right\rangle =\prod\limits_{p=1}^{\infty }\left(
u_{p}c_{+,p}^{\dag }+v_{p}c_{-,p+1}^{\dag }\right) \left\vert 0\right\rangle
\label{ask}
\end{equation}%
and the excitation energy for an antiskyrmion is%
\begin{equation}
\Delta _{aS}=E_{aS}-E_{GS}.  \label{dask}
\end{equation}%
In this case, the difference in angular momentum is $\Delta l_{z}=-\hslash $
and the rotation of the spins in real space is clockwise as shown in Fig. %
\ref{fig5}(b). The total number of reversed spin in the state $\left\vert
aS\right\rangle $ is again given by Eq. (\ref{flip}). At a given value of
the Zeeman coupling, $K$ is the same for skyrmion and antiskyrmion. In
total, the number of down spins in a skyrmion-antiskyrmion pair is given by $%
2K+1$ when counting the spin of the added electron and hole and $2K$ gives
the number of flipped spins. In an electron-hole excitation, $K=0.$

We remark that the skyrmion and antiskyrmion energies are modified by the
filled levels that we have neglected but the energy to create an unbound
skyrmion-antiskyrmion pair $\Delta _{S-aS}=\Delta _{aS}+\Delta _{S}$ is not.

When the Zeeman coupling is zero, the excitation energy of a large-scale
spin texture is given by the non-linear sigma model (NL$\sigma $M)%
\begin{equation}
E_{NL\sigma M}=\frac{1}{2}\rho _{s}\int \left( \nabla \mathbf{m}\right) ^{2},
\end{equation}%
where $\left\vert \mathbf{m}\right\vert =1$ is the spin field. In this
SU(2)-invariant limit, we know the exact spin stiffness which is given by%
\begin{equation}
\rho _{s}=\frac{1}{16\pi }\frac{e^{2}}{\kappa \ell }\int_{0}^{\infty
}dxx^{2}e^{-\frac{x^{2}}{2}}  \label{ros1}
\end{equation}%
in Landau level $n=0$ and by 
\begin{eqnarray}
\rho _{s} &=&\frac{1}{16\pi }\frac{e^{2}}{\kappa \ell }\int dxx^{2}e^{-\frac{%
x^{2}}{2}}  \label{ros2} \\
&&\times \frac{1}{4}\left[ L_{\left\vert n\right\vert }\left( \frac{x^{2}}{2}%
\right) +L_{\left\vert n\right\vert -1}\left( \frac{x^{2}}{2}\right) \right]
^{2}  \notag
\end{eqnarray}%
in other Landau levels. It follows that the exact energy of a single (large
scale) skyrmion or antiskyrmion is given by\cite{Rajaraman,Sondhi}.%
\begin{equation}
E_{NL\sigma M}=4\pi \rho _{s}.  \label{enlsm}
\end{equation}%
Eq. (\ref{enlsm}) is the energy needed to create a \textit{neutral} spin
texture. The definition of this energy\cite{MoonReview} is different from
the skyrmion excitation energy we introduced above. However, the energy to
create an unbound S-aS pair is given by $\Delta _{NL\sigma M}=$ $8\pi \rho
_{s}$ and this energy coincides\cite{Equasi} with $\Delta _{S-aS}$ as given
by Eqs. (\ref{dsk}) and (\ref{dask}).

The S-aS excitation gaps $\Delta _{NL\sigma M}$ for different values of $n$
have been computed by Kun Yang \textit{et al}.\cite{Kunyang} for a 2DEG with
Dirac bands and compared with the corresponding gaps for a 2DEG with
parabolic bands. This comparison showed that $\Delta _{NL\sigma M}<\Delta
_{e-h}$ for Landau levels $n=0,1,2,3$ so that the transport gap is dominated
by S-aS pairs at these filling factors.

In the $NL\sigma $ model, the energy of a skyrmion is independent of its
size and $K\rightarrow \infty .$ When the Zeeman coupling is considered,
skyrmions are smaller and it becomes necessary to consider the Hartree
electrostatic energy as well. The Hartree energy favors large scale
skyrmions while the Zeeman coupling favors small size skyrmions. The
competition between these energies lead to an optimal size for the skyrmion
at a given Zeeman coupling.

Fig. \ref{fig8} shows the transports gaps $\Delta _{e-h}$ and $\Delta
_{NL\sigma M}$ for several Landau levels at $\Delta _{Z}=0.$ The transport
gaps for the chiral and non-chiral 2DEG's are also listed in Table 1 of Ref. %
\onlinecite{Kunyang}.

\section{GREEN'S FUNCTION FORMALISM FOR SKYRMIONS}

In the skyrmion and antiskyrmion states, the only non-zero $\left\langle
\rho \right\rangle ^{\prime }s$ are given by 
\begin{equation}
\left\langle \rho _{p,p}^{s,s}\right\rangle ,\left\langle \rho _{p\pm
1,p}^{-,+}\right\rangle ,\left\langle \rho _{p,p\pm 1}^{+,-}\right\rangle
\neq 0,
\end{equation}%
where the upper(lower) sign in the subscripts is for skyrmion(antiskyrmion).
To compute these average values, we define the matrix of Matsubara Green's
functions%
\begin{equation}
G_{p}^{\pm }\left( \tau \right) =\left( 
\begin{array}{cc}
G_{p,p}^{+,+}\left( \tau \right) & G_{p,p\pm 1}^{+,-}\left( \tau \right) \\ 
G_{p\pm 1,p}^{-,+}\left( \tau \right) & G_{p\pm 1,p\pm 1}^{-,-}\left( \tau
\right)%
\end{array}%
\right) ,
\end{equation}%
with%
\begin{equation}
G_{p,p^{\prime }}^{s,s^{\prime }}\left( \tau \right) =-\left\langle T_{\tau
}c_{s,p}\left( \tau \right) c_{s^{\prime },p^{\prime }}^{\dag }\left(
0\right) \right\rangle ,
\end{equation}%
where $T_{\tau }$ is the imaginary time ordering operator. By definition,
the $\left\langle \rho \right\rangle ^{\prime }s$ are related to the Green's
functions by the relation 
\begin{equation}
G_{p}^{\pm }\left( \tau =0^{-}\right) =\left( 
\begin{array}{cc}
\left\langle \rho _{p,p}^{+,+}\right\rangle & \left\langle \rho _{p\pm
1,p}^{-,+}\right\rangle \\ 
\left\langle \rho _{p,p\pm 1}^{+,-}\right\rangle & \left\langle \rho _{p\pm
1,p\pm 1}^{-,-}\right\rangle%
\end{array}%
\right) .
\end{equation}%
Note that we must take $p=0,1,2,...$for a skyrmion and $p=1,2,3,...$ for an
antiskyrmion. To take into account the added electron or hole, we must in
addition set $\left\langle \rho _{0,0}^{-,-}\right\rangle =1$ for the
skyrmion and $\left\langle \rho _{0,0}^{+,+}\right\rangle =0$ for the
antiskyrmion.

Using the Heisenberg equation of motion 
\begin{equation}
\hslash \frac{\partial }{\partial \tau }\left( \ldots \right) =\left[
K,\left( \ldots \right) \right] ,
\end{equation}%
with $K=H-\mu N_{e},$ where $\mu $ is the chemical potential, we get the
following equation of motion for the Green's function $G_{p,p^{\prime
}}^{s,s^{\prime }}\left( \tau \right) $:%
\begin{eqnarray}
\hslash \frac{\partial }{\partial \tau }G_{p,p^{\prime }}^{s,s^{\prime
}}\left( \tau \right) &=&-\hslash \delta \left( \tau \right) \delta
_{p,p^{\prime }}\delta _{s,s^{\prime }}-E_{s}^{0}G_{p,p^{\prime
}}^{s,s^{\prime }}\left( \tau \right) \\
&&-\sum_{s^{\prime \prime }}\sum_{m}V_{m,m,p,p}^{n}\left\langle \rho
_{m,m}^{s^{\prime \prime },s^{\prime \prime }}\right\rangle G_{p,p^{\prime
}}^{s,s^{\prime }}\left( \tau \right)  \notag \\
&&+\sum_{m}V_{m,p,p,m}^{n}\left\langle \rho _{m,m}^{s,s}\right\rangle
G_{p,p^{\prime }}^{s,s^{\prime }}\left( \tau \right)  \notag \\
&&+\sum_{m}V_{m\pm 1,p\pm 1,p,m}^{n}\left\langle \rho _{m\pm
1,m}^{-,+}\right\rangle G_{p\pm 1,p^{\prime }}^{-,s^{\prime }}\left( \tau
\right) \delta _{s,+}  \notag \\
&&+\sum_{m}V_{m,p\mp 1,p,m\pm 1}^{n}\left\langle \rho
_{m,m+1}^{+,-}\right\rangle G_{p\mp 1,p^{\prime }}^{+,s^{\prime }}\left(
\tau \right) \delta _{s,-}  \notag \\
&&+\sum_{m}V_{m,m,p,p}^{n}G_{p,p^{\prime }}^{s,s^{\prime }}\left( \tau
\right) .  \notag
\end{eqnarray}%
This equation can be written in an obvious matrix form as%
\begin{equation}
\left[ \left( i\omega _{n}+\mu \right) I-\frac{1}{\hslash }F_{p}^{\pm }%
\right] G_{p}^{\pm }\left( i\omega _{n}\right) =I,  \label{w11}
\end{equation}%
where $I$ is the $2\times 2$ units matrix. The components of the $2\times 2$
matrices $F_{p}^{\pm }$ are given by (with $\left\langle \rho
_{m,m}\right\rangle \equiv \sum_{s}\left\langle \rho
_{m,m}^{s,s}\right\rangle $) 
\begin{eqnarray}
\left( F_{p}^{\pm }\right) _{1,1} &=&E_{+}^{0}+\sum_{m}V_{m,m,p,p}^{n}\left[
\left\langle \rho _{m,m}\right\rangle -1\right] \\
&&-\sum_{m}V_{m,p,p,m}^{n}\left\langle \rho _{m,m}^{+,+}\right\rangle , 
\notag
\end{eqnarray}%
\begin{eqnarray}
\left( F_{p}^{\pm }\right) _{2,2} &=&E_{-}^{0}+\sum_{m}V_{m,m,p\pm 1,p\pm
1}^{n}\left[ \left\langle \rho _{m,m}\right\rangle -1\right] \\
&&-\sum_{m}V_{m,p\pm 1,p\pm 1,m}^{n}\left\langle \rho
_{m,m}^{-,-}\right\rangle ,  \notag
\end{eqnarray}%
\begin{equation}
\left( F_{p}^{\pm }\right) _{1,2}=-\sum_{m}V_{m\pm 1,p\pm
1,p,m}^{n}\left\langle \rho _{m\pm 1,m}^{-,+}\right\rangle ,
\end{equation}%
\begin{equation}
\left( F_{p}^{\pm }\right) _{2,1}=-\sum_{m}V_{m,p,p\pm 1,m\pm
1}^{n}\left\langle \rho _{m,m\pm 1}^{+,-}\right\rangle ,
\end{equation}%
where the summations extend over all values of $m$ in the diagonal elements
of $F_{p}^{\pm }$ and from $m=0\left( m=1\right) $ to infinity for skyrmions
(antiskyrmions) in the off-diagonal elements.

The matrices $F^{\pm }$ are Hermitian and can be diagonalized by a unitary
transformation 
\begin{equation}
F_{p}^{\pm }=U_{p}^{\pm }D_{p}^{\pm }\left( U_{p}^{\pm }\right) ^{\dag },
\end{equation}%
with $D_{p}^{\pm }$ the diagonal matrix of the eigenvalues $\left(
d_{p}^{\pm }\right) _{k}$ of $F_{p}^{\pm }.$ If follows that 
\begin{equation}
\left[ G_{p}^{\pm }\left( i\omega _{n}\right) \right] _{i,j}=\sum_{k}\frac{%
\left( U_{p}^{\pm }\right) _{i,k}\left[ \left( U_{p}^{\pm }\right) \right]
_{k,j}^{\dag }}{i\omega _{n}+\mu -\left( d_{p}^{\pm }\right) _{k}}
\end{equation}%
and%
\begin{equation}
\left[ G_{p}^{\pm }\left( \tau =0^{-}\right) \right] _{i,j}=\left(
U_{p}^{\pm }\right) _{i,n}\left[ \left( U_{p}^{\pm }\right) ^{\dag }\right]
_{n,j},
\end{equation}%
with $n=1$ when $\left( d_{p}^{\pm }\right) _{1}<\left( d_{p}^{\pm }\right)
_{2}$ and $n=2$ otherwise.

The Hartree-Fock self-consistent Eq. (\ref{w11}) is solved numerically using
an iterative scheme until self-consistency is achieved for the $\left\langle
\rho \right\rangle ^{\prime }s.$ The excitation energy for skyrmion (upper
sign) and antiskyrmion (lower sign) are then computed using (the Zeeman term
is absent for valley skyrmions):

\begin{eqnarray}
\Delta _{S/aS} &=&\left\langle H_{\pm }\right\rangle -E_{GS} \\
&=&-\frac{1}{2}g\mu _{B}B\sum_{s,m}s\left[ \left\langle \rho
_{m,m}^{s,s}\right\rangle -\delta _{s,+}\right]  \notag \\
&&+\frac{1}{2}\sum_{m,p}V_{m,m,p,p}^{n}\left[ \left( \left\langle \rho
_{m,m}\right\rangle -2\right) \left\langle \rho _{p,p}\right\rangle +1\right]
\notag \\
&&-\frac{1}{2}\sum_{s}\sum_{m,p}V_{m,p,p,m}^{n}\left[ \left\langle \rho
_{m,m}^{s,s}\right\rangle \left\langle \rho _{p,p}^{s,s}\right\rangle
-\delta _{s,+}\right]  \notag \\
&&-\frac{1}{2}\sum_{m,p}V_{m,p,p\pm 1,m\pm 1}^{n}\left\langle \rho _{m,m\pm
1}^{+,-}\right\rangle \left\langle \rho _{p\pm 1,p}^{-,+}\right\rangle 
\notag \\
&&-\frac{1}{2}\sum_{m,p}V_{m\pm 1,p\pm 1,p,m}^{n}\left\langle \rho _{m\pm
1,m}^{-,+}\right\rangle \left\langle \rho _{p,p\pm 1}^{+,-}\right\rangle . 
\notag
\end{eqnarray}%
The excitation energy combines the Zeeman cost of the flipped spins, the
Coulomb self-interaction of the excess charge and the exchange energy cost
associated the rotation of the spins with respect to the ferromagnetic
ground state.

The change in the electronic density and spin pattern in real space can
easily be obtained from the $\left\langle \rho \right\rangle ^{\prime }s$ by
using%
\begin{equation}
\delta n\left( \mathbf{r}\right) =\sum_{m}\Lambda _{m}^{n}\left( \mathbf{r}%
\right) \left[ \left\langle \rho _{m,m}^{+,+}\right\rangle +\left\langle
\rho _{m,m}^{-,-}\right\rangle -1\right] ,  \label{deltannn}
\end{equation}%
and%
\begin{eqnarray}
\delta S_{x}\left( \mathbf{r}\right) &=&\frac{\hslash }{2}\sum_{m}\left[
\Upsilon _{m,\pm }^{n}\left( \mathbf{r}\right) \left\langle \rho _{m,m\pm
1}^{+,-}\right\rangle +c.c.\right] , \\
\delta S_{y}\left( \mathbf{r}\right) &=&\frac{\hslash }{2i}\sum_{m}\left[
\Upsilon _{m,\pm }^{n}\left( \mathbf{r}\right) \left\langle \rho _{m,m\pm
1}^{+,-}\right\rangle -c.c.\right] , \\
\delta S_{z}\left( \mathbf{r}\right) &=&\frac{\hslash }{2}\sum_{m}\Lambda
_{m}^{n}\left( \mathbf{r}\right) \left[ \left\langle \rho
_{m,m}^{+,+}\right\rangle -\left\langle \rho _{m,m}^{-,-}\right\rangle -1%
\right] ,
\end{eqnarray}%
with the definitions%
\begin{eqnarray}
\Lambda _{m}^{0}\left( \mathbf{r}\right) &=&\left\vert \varphi _{0,m}\left( 
\mathbf{r}\right) \right\vert ^{2}, \\
\Upsilon _{m,\pm }^{0}\left( \mathbf{r}\right) &=&\varphi _{0,m}^{\ast
}\left( \mathbf{r}\right) \varphi _{0,m\pm 1}\left( \mathbf{r}\right) ,
\end{eqnarray}%
and%
\begin{eqnarray}
\Lambda _{m}^{\left\vert n\right\vert >0}\left( \mathbf{r}\right) &=&\frac{1%
}{2}\left[ \left\vert \varphi _{\left\vert n\right\vert ,m}\left( \mathbf{r}%
\right) \right\vert ^{2}+\left\vert \varphi _{\left\vert n\right\vert
-1,m}\left( \mathbf{r}\right) \right\vert ^{2}\right] , \\
\Upsilon _{m,\pm }^{\left\vert n\right\vert >0}\left( \mathbf{r}\right) &=&%
\frac{1}{2}\varphi _{\left\vert n\right\vert ,m}^{\ast }\left( \mathbf{r}%
\right) \varphi _{\left\vert n\right\vert ,m\pm 1}\left( \mathbf{r}\right) \\
&&+\frac{1}{2}\varphi _{\left\vert n\right\vert -1,m}^{\ast }\left( \mathbf{r%
}\right) \varphi _{\left\vert n\right\vert -1,m\pm 1}\left( \mathbf{r}%
\right) .  \notag
\end{eqnarray}%
Again, the summations over $m$ in $\delta S_{x}$ and $\delta S_{y}$ run from 
$m=0\left( m=1\right) $ to infinity for skyrmions(antiskyrmions).

\section{NUMERICAL\ RESULTS FOR THE EXCITATION ENERGIES}

The Green's function approach just described is well-suited to compute the
energy of a finite-size skyrmion but there is a practical difficulty with
it. In practice we are forced to truncate the set of single-particle angular
momenta that we include in the description of both the ferromagnetic ground
state and the charged excitations at a finite value $m_{\max }.$ Since the
single-particle orbital with angular momentum $m$ is localized near a ring
with radius $\sqrt{\left( 2m+1\right) }\ell ,$ this is equivalent to working
with a finite-size electron disk of radius $R\approx \sqrt{2m_{\max }}\ell .$
The skyrmion excitation energy $\Delta _{S}$ will be given accurately by our
method if the tail of the disturbance associated with the charged excitation
does not extend to the edge of the disk. When the Zeeman coupling $%
\widetilde{g}\equiv \Delta _{Z}/\left( e^{2}/\kappa \ell \right) $ decreases
below a certain value, the skyrmion size becomes large and this condition is
not satisfied.

The Landau level wave functions obey the identity%
\begin{equation}
\sum_{m=0}^{m_{\max }=\infty }\left\vert \varphi _{n,m}\left( \mathbf{r}%
\right) \right\vert ^{2}=\frac{1}{2\pi \ell ^{2}}.  \label{identity}
\end{equation}%
If $m_{\max }=160,$ Eq. (\ref{identity}) is satisfied numerically for $%
r_{\max }/\ell \lesssim 15$ while for $m_{\max }=1000,$ it is satisfied for $%
r/\ell \lesssim 35.$ In our numerical calculations, we set $m_{\max }=1000$.
It follows that the charged excitation that we compute must be well
contained in a disk a radius $r_{\max }/\ell \lesssim 35$ for our
calculation to be reliable.

\begin{figure}[tbph]
\includegraphics[scale=1.0]{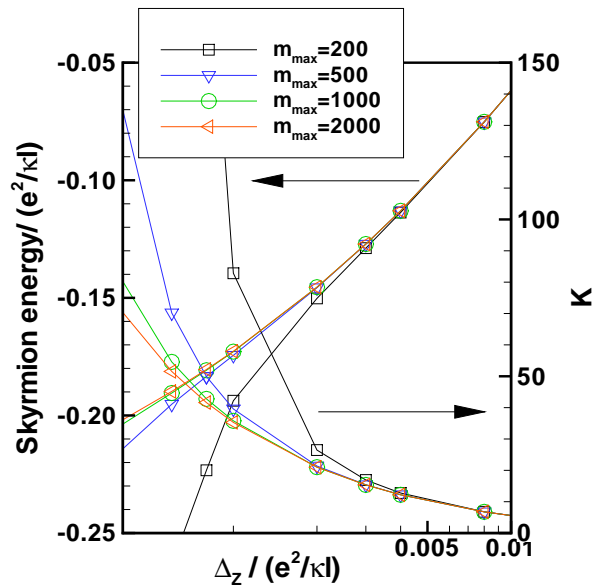}
\caption{(Color online) Energy of one skyrmion $E_{S}$ and the corresponding
number of down spins $N_{\downarrow }=2K+1$ as a function of the Zeeman
coupling for different values of the maximum angular momentum used in the
computation.}
\label{fig1}
\end{figure}

Fig. \ref{fig1} shows the energy $E_{S}$ of one skyrmion and the
corresponding number of down spins $N_{\downarrow }=2K+1$ as a function of
the Zeeman coupling $\widetilde{g}$ for different values of the maximum
angular momentum $m_{\max }$ up to $2000$ (In all our numerical
calculations, we use $\kappa =2.5$ for the dielectric constant of the
substrate.) A good convergence of $E_{S}$ and $N_{\downarrow }$ is obtained
for $\widetilde{g}\gtrsim 0.001$ with $m_{\max }=1000.$ Due to the
variational nature of the Hartree-Fock calculation, the energies approach
their asymptotic value much more rapidly with increasing $m_{\max }$ than
estimates of the optimal value of $N_{\downarrow }.$ The values of $\Delta
_{S-aS}$ at small Zeeman coupling are thus more reliable than those of $%
N_{\downarrow }.$

\begin{figure}[tbph]
\includegraphics[scale=1.0]{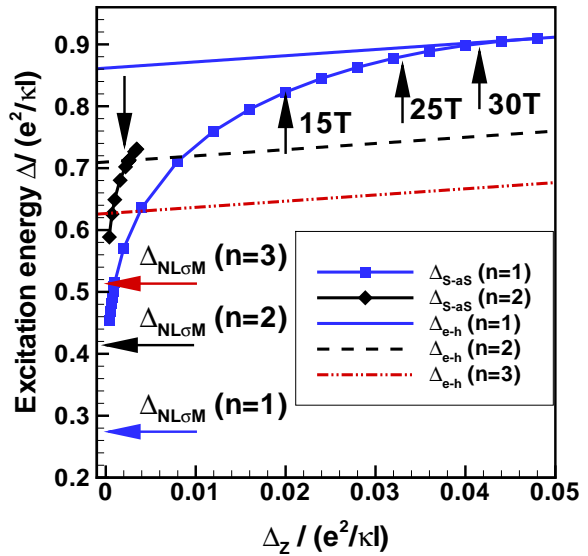}
\caption{(Color online) Excitation energy of a skyrmion-antiskyrmion pair $%
\Delta _{SK-ASK}$ and an electron-hole pair $\Delta _{e-h}$ as a function of
the Zeeman coupling $\Delta _{Z}/(e^{2}/\protect\kappa \ell $) for Landau
levels $n=1,2,3$. The horizontal arrows indicate the value of the
skyrmion-antiskyrmion gap $\Delta _{NL\protect\sigma M}$ calculated in the
non-linear $\protect\sigma $ model. The upward vertical arrows are
positioned at the value of $\protect\widetilde{g}$ corresponding to total
magnetic fields $B=15,25,30$ T when $B_{\bot }=15$ T. The downward arrow
points to the value of $\protect\widetilde{g}_{c}$ for $n=2$.}
\label{fig2}
\end{figure}

Fig. \ref{fig2} shows the behavior of the gaps $\Delta _{e-h}$ and $\Delta
_{S-aS}$ with Zeeman coupling $\widetilde{g}$ for Landau levels $n=1,2,3.$
The value of the gap $\Delta _{NL\sigma M}$ is also indicated for each
Landau level. The upward vertical arrows are placed at values of $\widetilde{%
g}$ corresponding to the total magnetic fields $B=15,25,30$ T with $B_{\bot
}=15$ T. This allows a comparison of our results with Fig. 2(e) of Ref. %
\onlinecite{Afyoung} where the activation gap measured in a tilted-field
experiment is plotted as a function of the total magnetic field for $n=-1,-2$%
. (Note that the theoretical activation energies depend only on $\left\vert
n\right\vert .$) In such experiments, the magnetic length is actually
defined by $\ell =\sqrt{\hslash c/eB_{\bot }}$ and $B_{\bot }$ is kept fixed
while the magnetic field is tilted. This is equivalent to changing $%
\widetilde{g}$ and keeping the filling factor fixed.

In Fig. \ref{fig2}, the gap $\Delta _{e-h}$ \textit{decreases} with
increasing Landau level index reflecting the decrease of the exchange energy
with $n$ in Eq. (\ref{deltaeh}). On the contrary, $\Delta _{NL\sigma M}$ and 
$\Delta _{S-aS}$ both \textit{increase} with $n$ in the small Zeeman range
where skyrmions exist for $n=1$ and $n=2$ (the region near the down arrow in
the figure). If the $n=2$ skyrmion were to persist to larger values of $%
\widetilde{g},$ the $\Delta _{S-aS}$ gap would actually decrease with $n$ at
large $\widetilde{g}$ but this does not happen in our calculation.

\begin{figure}[tbph]
\includegraphics[scale=1.0]{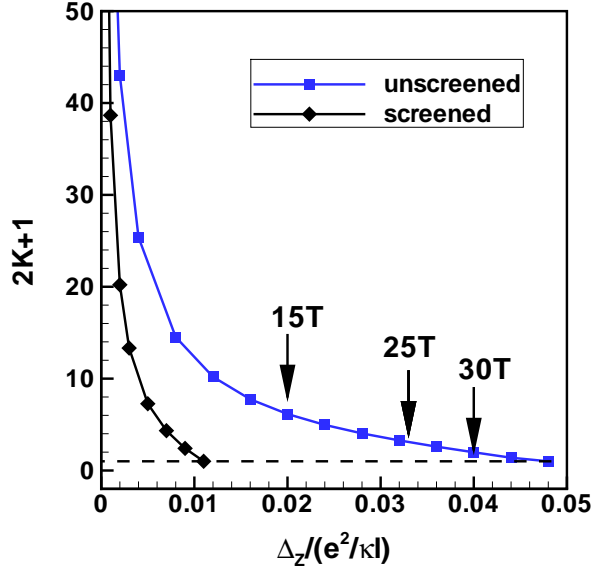}
\caption{(Color online) Number of down spins $N_{\downarrow }=2K+1$ in a
skyrmion-antiskyrmion pair as a function of the Zeeman coupling $\Delta
_{Z}/\left( e^{2}/\protect\kappa \ell \right) $ for Landau level $n=1$ with
and without screening corrections. The arrows are placed that the values of $%
\protect\widetilde{g}$ where the total magnetic fields $B=15,25,30$ T when
the perpendicular component $B_{\bot }=15 $ T. The dashed line indicates $%
N_{\downarrow }=1,$ the electron-hole limit.}
\label{fig3}
\end{figure}

The maximal value, $\widetilde{g}_{c},$ of the Zeeman coupling for which $%
\Delta _{S-aS}<\Delta _{e-h}$ decreases dramatically with Landau level level
index as shown in Fig. \ref{fig2}. For example, the value of $\widetilde{g}%
_{c}\approx 0.0026$ for $n=2$ is one order of magnitude lower than that for $%
n=1.$ This makes skyrmions difficult to calculate in higher Landau levels.
Interestingly, we find that for $n=2,$ $\Delta _{S-aS}$ goes over $\Delta
_{e-h}$ for $g\geq $ $\widetilde{g}_{c}$ ($\widetilde{g}_{c}$ is indicated
by a downward arrow in Fig. \ref{fig2}) instead of reaching $\Delta _{e-h}$
smoothly as is the case for $n=1.$ At the crossing point $\widetilde{g}_{c}$
for $n=2$, the number of down spins, $N_{\downarrow }$ in the S-aS pair (see
Fig. \ref{fig3}) is $\approx 25,$ a large value. A similar jump in the
number of down spins at the transition from S-aS pair to electron-hole pair
was predicted theoretically for skyrmions in a conventional 2DEG when the
finite width of the well was taken into account and at filling factor $\nu
=3 $\cite{Fertig2}. This jumps suggests that the spin polarization of the
C2DEG could change abruptly at $\widetilde{g}_{c}.$ This first order
transition has been seen experimentally in a conventional 2DEG at $\nu =1$%
\cite{Melinte} and also in a conventional bilayer 2DEG at $\nu =1$ when the
electrons occupy only one of the two layers\cite{Sawada}. Our calculation
shows that it can also happen in graphene.

The number of down spins $N_{\downarrow }>1$ for a S-aS pair while $%
N_{\downarrow }=1$ for an electron hole pair. Fig. \ref{fig3} shows that the
rapid increase in energy of $\Delta _{S-aS}$ with $\widetilde{g}$ is
associated with a rapid decrease in $N_{\downarrow }.$ The number of down
spins varies roughly linearly with $\widetilde{g}$ in between $B_{\bot }=15$
T and $B_{\bot }=30$ T but not at smaller values of the Zeeman coupling. At $%
B_{\bot }=15$ T, $N_{\downarrow }\approx 6$ for $n=1$ corresponding to $%
K=2.5 $ reversed spins per skyrmion. Fig. \ref{fig3} shows that, for the
same Zeeman coupling, $K$ is smaller for a $n=2$ than for a $n=1$ skyrmion.

\begin{figure}[tbph]
\includegraphics[scale=1.0]{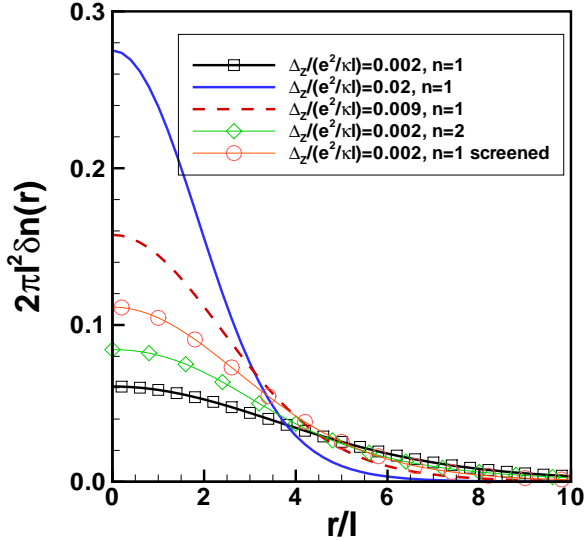}
\caption{(Color online)Profile of the induced density $\protect\delta %
n\left( r\right) $ when a skyrmion is added to the ground state in Landau
level $n=1$ for several values of the Zeeman coupling $\Delta _{Z}/\left(
e^{2}/\protect\kappa \ell \right) $. The profiles for the screened skyrmion
in $n=1$ and the unscreened skyrmion in $n=2$ are also shown. }
\label{fig4}
\end{figure}

Fig. \ref{fig4} shows $\delta n\left( r\right) ,$ the change in the density
of the C2DEG with respect to the ferromagnetic ground state density $%
n_{GS}\left( r\right) =1/2\pi \ell ^{2}$ when a skyrmion is added to the
ground state. (Because of the electron-hole symmetry of the Hamiltonian near
half-filling$,$ $\delta n_{S}\left( r\right) \mathbf{=-}\delta n_{aS}\left(
r\right) $). We can define the size or radius of a skyrmion, $r_{sky},$ by
the condition $\delta n\left( r=r_{sky}\right) /\delta n\left( r=0\right)
=1/2$. Fig. \ref{fig4} shows that the size of the skyrmions shrinks with
increasing Zeeman coupling and also with increasing Landau level index at
fixed Zeeman coupling. For $\widetilde{g}=0.002,$ the skyrmion size for $n=1$
is $r_{sky}/\ell \approx 2$ and the tail of the $\delta n\left( r\right) $
is well within the maximal radius $r_{\max }/\ell =35$ discussed above.

For $n=3,$ the crossing point $\widetilde{g}_{c}$ occurs at a value of $%
\widetilde{g}\lesssim 0.0002$ where $N_{\downarrow }$ is very large,
suggesting that the skyrmion size at that Zeeman coupling is already beyond
the limit of reliability of our approach. Since the NL$\sigma $M result
indicates that skyrmions are the lowest-energy charged excitations for $n=3,$
we can conclude that, if they persist to finite Zeeman coupling, it is
certainly in a very narrow range of $\widetilde{g},$ approximately an order
of magnitude smaller than for $n=2.$

\begin{figure}[tbph]
\includegraphics[scale=1.0]{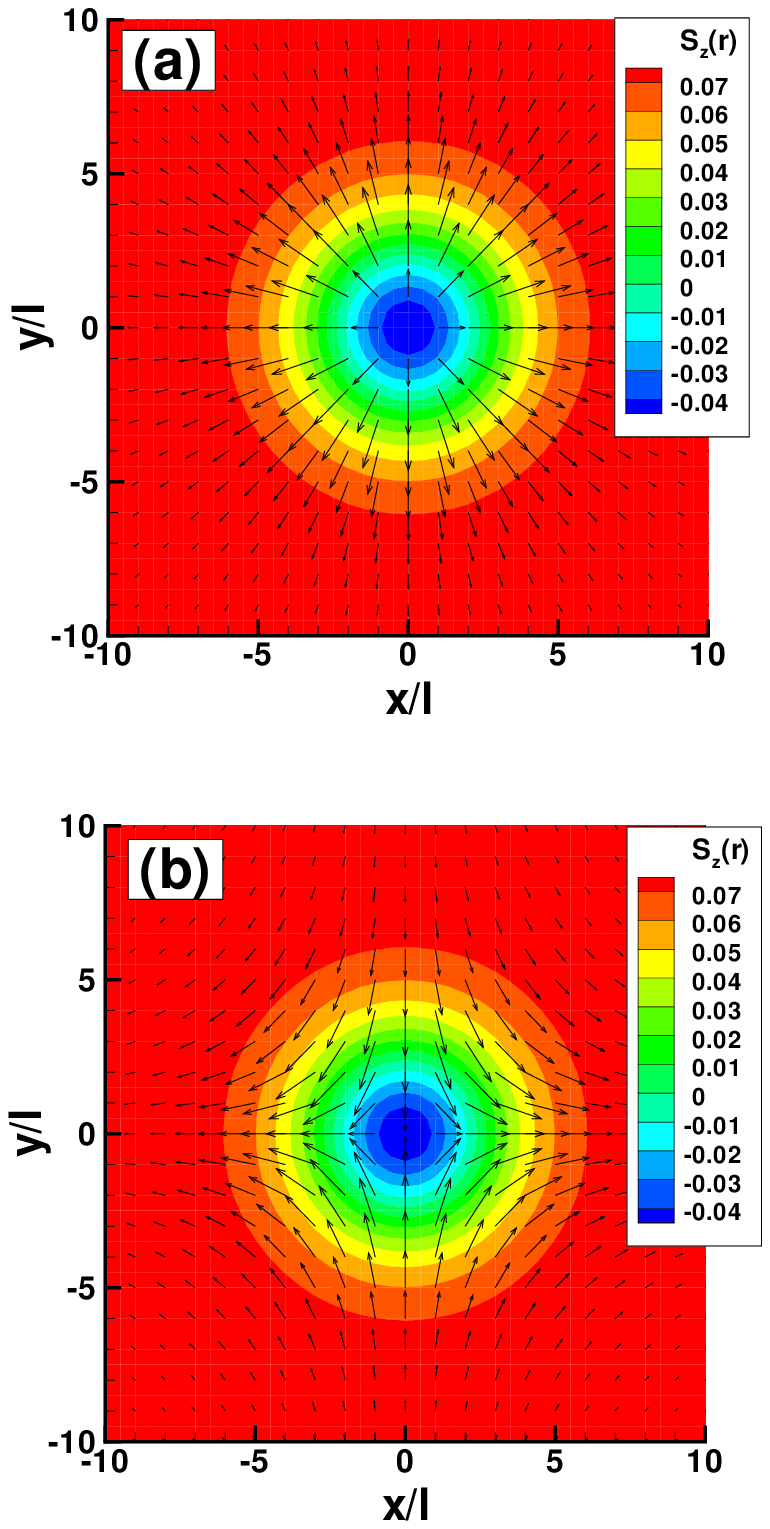}
\caption{(Color online)Spin texture in the $x-y$ plane for a (a) skyrmion
and (b) an antiskyrmion in Landau level $n=1$ at Zeeman coupling $\Delta
_{Z}/\left( e^{2}/\protect\kappa \ell \right) =0.011.$ The density plot
shows $S_{z}\left( r\right) $ in units of $\hslash /(2\protect\pi \ell ^{2}).
$}
\label{fig5}
\end{figure}

The spin texture $\mathbf{S}_{\Vert }\left( \mathbf{r}\right) $ for a
skyrmion and an antiskyrmion excitations in $n=1$ at $\widetilde{g}=0.011$
is plotted in Fig. \ref{fig5} with the component $S_{z}\left( \mathbf{r}%
\right) $ given by the superimposed density plot. The in-plane component of
the spin makes a $2\pi $ counterclockwise (skyrmion) or clockwise
(antiskyrmion) rotation around the center of the topological charge.

Finally, the magnitude of the gap $\Delta _{S-aS}\approx e^{2}/\kappa \ell .$
For $B_{\bot }=15$ T, $e^{2}/\kappa \ell =1011$ K and so $\Delta
_{S-aS}\approx 910$ K at $B=30$ T for $n=1$.

Our results can be compared with those of Ref. \onlinecite{Afyoung} (see
Fig. 2(e) of this paper) where the transport gap was measured at total
magnetic fields $B=15,25,30$ T with the perpendicular magnetic field $%
B_{\bot }=15$ T kept fixed. In this experiment, the electronic density was
varied in order to study the spin-texture excitations at filling factors $%
\nu =-4,-8,-12$ corresponding to half-filling of Landau levels $n=-1,-2,-3.$
At $B=15,25,30$ T, according to our calculations, the transport gap is given
by $\Delta _{S-aS}^{\left( n=1\right) }$ in $n=1$ and by $\Delta
_{e-h}^{\left( n=2\right) },\Delta _{e-h}^{\left( n=3\right) }$ in $n=2$ and 
$n=3$ with the ordering $\Delta _{S-aS}^{\left( n=1\right) }>\Delta
_{e-h}^{\left( n=2\right) }>\Delta _{e-h}^{\left( n=3\right) }.$ This
ordering is consistent with the experimental result except for $n=2.$ In
this case, the experiment measures a small number of spin flips i.e. $%
N_{\downarrow }\approx 1.4$ suggesting that $\Delta _{S-aS}^{\left(
n=2\right) }<\Delta _{e-h}^{\left( n=2\right) }.$ We cannot explain this
difference with our model of skyrmion excitations.

Another difference between the experimental and theoretical results is the
size of the transport gap. For example, the experimental value of $\Delta
_{S-aS}^{\left( n=1\right) }\approx 75$ K at $B=30$ T while we find $\Delta
_{S-aS}\approx 910$ K, a much larger value. Several effects may affect our
results such as disorder, Landau level mixing\cite{Fertig} and screening. In
a conventional 2DEG, taking into account the quantum well width\cite{Fertig2}
is known to decrease the excitation energy but this effect is not present in
graphene. In the remainder of this paper, we study the corrections due to
screening since they are easy to include in our calculation and they lead to
a substantial decrease of the gap. We leave disorder and Landau mixing
effects to further work. Valley skyrmions were also studied at filling
factors $\nu =-3,-5$ in Ref. \onlinecite{Afyoung}. We comment on them in the
next section.

In closing this section, we remark that we have shown and commented here our
results for the excitation energy of a skyrmion-antiskyrmion pair.
Nevertheless, we have verified that whenever this energy is smaller than the
corresponding electron-hole pair energy, the skyrmion (antiskyrmion) energy
is smaller than the electron(hole) energy.

\section{SCREENING\ CORRECTIONS TO THE EXCITATION GAPS}

To include screening, we follow the approach of Ref. \onlinecite{Aleiner}
where it was shown that when the Landau levels other than the partially
filled level are integrated out, the low-frequency dynamics of the 2DEG is
described by the electrons belonging to the partially filled Landau level
but the interaction between these electrons (and with the positive charge of
the background) is renormalized due to the polarizability of all the other
Landau levels. In the Hartree-Fock approximation, this renormalization
amounts to screen both the Hartree and Fock interactions\cite{Note3}. The
bare Coulomb interaction $V\left( q\right) =2\pi e^{2}/\kappa q$ must then
be replaced by $V\left( q\right) =2\pi e^{2}/\varepsilon \left( q\right)
\kappa q$ where $\varepsilon \left( q\right) $ is the static dielectric
function calculated in the random-phase approximation (RPA). Such procedure
was used, for example, in the study of inhomogeneous states such as bubble
and stripe phases in quantum Hall systems\cite{Fogler}.

We follow this procedure by using the screened Coulomb interaction in the
matrix elements of the Coulomb interaction in Eq. (\ref{elements}). The
matrix elements are recalculated by inserting the dielectric function i.e. $%
\left( \frac{e^{2}}{\kappa \ell }\right) \int_{0}^{\infty
}dxe^{-2x^{2}}\left( \ldots \right) \rightarrow \left( \frac{e^{2}}{\kappa
\ell }\right) \int_{0}^{\infty }dx\frac{e^{-2x^{2}}}{\varepsilon \left(
x\right) }\left( \ldots \right) $ where $x=q\ell /\sqrt{2}$ in Eq. (\ref%
{vm1m2}). The dielectric function is evaluated in the random-phase
approximation and is given by

\begin{equation}
\varepsilon \left( \mathbf{q}\right) =1-\frac{2\pi e^{2}}{q}\chi
^{0,R}\left( \mathbf{q},\omega =0\right) ,
\end{equation}%
where $\chi ^{0,R}\left( \mathbf{q,}\omega \right) $ is the retarded density
response function computed for a non-interacting C2DEG in a magnetic field.
More precisely,%
\begin{eqnarray}
\varepsilon \left( \mathbf{q}\right) &=&1+\frac{e^{2}/\kappa \ell }{\hslash
\omega _{c}^{\ast }}\frac{1}{q\ell }\sum_{\alpha ,s}\sum_{n,n^{\prime
}}\left\vert \Xi _{n,n^{\prime }}\left( \mathbf{q}\right) \right\vert ^{2}
\label{epsilon} \\
&&\times \frac{\nu _{n,\alpha ,s}-\nu _{n^{\prime },\alpha ,s}}{sgn\left(
n^{\prime }\right) \sqrt{\left\vert n^{\prime }\right\vert }-sgn\left(
n\right) \sqrt{\left\vert n\right\vert }},  \notag
\end{eqnarray}%
where $\nu _{n,\alpha ,s}$ is the filling factor of Landau level $n$ with
valley index $\alpha $ and spin $s$ and the function%
\begin{eqnarray}
\Xi _{n,n^{\prime }}\left( \mathbf{q}\right) &=&\frac{1}{2}\Theta \left(
\left\vert n\right\vert \right) \Theta \left( \left\vert n^{\prime
}\right\vert \right) \\
&&\times \left[ F_{\left\vert n\right\vert ,\left\vert n^{\prime
}\right\vert }\left( \mathbf{q}\right) +sgn\left( n\right) sgn\left(
n^{\prime }\right) F_{\left\vert n\right\vert -1,\left\vert n^{\prime
}\right\vert -1}\left( \mathbf{q}\right) \right]  \notag \\
&&+\frac{1}{\sqrt{2}}\left[ \delta _{n,0}\Theta \left( \left\vert n^{\prime
}\right\vert \right) +\delta _{n^{\prime },0}\Theta \left( \left\vert
n\right\vert \right) \right] F_{\left\vert n\right\vert ,\left\vert
n^{\prime }\right\vert }\left( \mathbf{q}\right)  \notag \\
&&+\delta _{n,0}\delta _{n^{\prime },0}F_{0,0}\left( \mathbf{q}\right) , 
\notag
\end{eqnarray}%
with%
\begin{eqnarray}
F_{n,n^{\prime }}\left( \mathbf{q}\right) &=&\left( \frac{n^{\prime }!}{n!}%
\right) ^{1/2}\left[ \frac{\left( q_{y}+iq_{x}\right) \ell }{\sqrt{2}}\right]
^{n-n^{\prime }} \\
&&\times L_{n^{\prime }}^{n-n^{\prime }}\left( \frac{q^{2}\ell ^{2}}{2}%
\right) e^{-q^{2}\ell ^{\ell 2}/4},  \notag
\end{eqnarray}%
for $n\geq n^{\prime }.$ For $n<n^{\prime }$ $F_{n,n^{\prime }}\left( 
\mathbf{q}\right) =\left[ F_{n^{\prime },n}\left( -\mathbf{q}\right) \right]
^{\ast }$. We have defined the effective cyclotron energy $\hslash \omega
_{c}^{\ast }=\sqrt{2}\hslash v_{F}/\ell $ so that%
\begin{equation}
\frac{e^{2}/\kappa \ell }{\hslash \omega _{c}^{\ast }}=\frac{1}{\sqrt{2}%
\kappa }\alpha ^{\ast }
\end{equation}%
where $\alpha ^{\ast }=e^{2}/\hslash v_{F}$ is the effective fine structure
constant for graphene. The dielectric function at integer filling $\nu $ of
electrons is equal to the dielectric function at integer filling $\nu $ of
holes. We write this as $\varepsilon _{\nu }=\varepsilon _{-\nu }.$

\begin{figure}[tbph]
\includegraphics[scale=1.0]{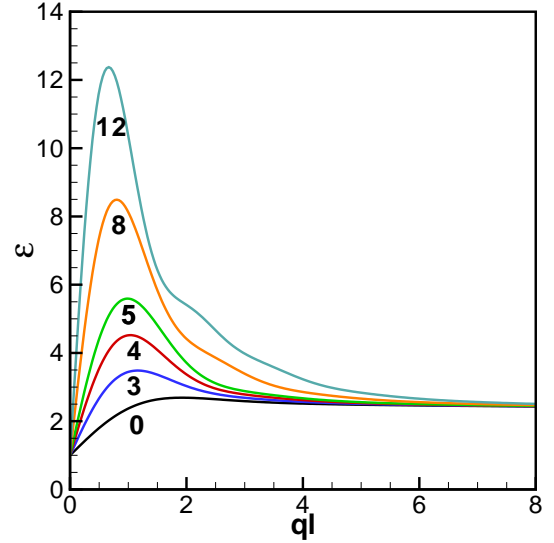}
\caption{(Color online)Static dielectric function computed in the
random-phase approximation at different filling factors $\left\vert \protect%
\nu \right\vert $ (indicated by the number below each curve) in Landau
levels $\left\vert n\right\vert =0,1,2,3.$}
\label{fig6}
\end{figure}

The dielectric function $\varepsilon \left( \mathbf{q}\right) $ of the C2DEG
has been evaluated previously\cite{Screening}. We show our results obtained
at different filling factors in Fig. \ref{fig6}$.$ In our calculation we
include Landau levels $m\in \left[ -800,800\right] $ in the summations in
Eq. (\ref{epsilon}). The dielectric function $\varepsilon \left( q\right) =1$
at $q=0$ and $q\rightarrow \infty .$ It is maximal around $q\ell \approx 1$
and increases with increasing filling factors$.$ In particular, screening is
larger at $3/4$ filling of a given Landau level than at $1/4$ filling as
shown in Fig. \ref{fig6} for $\left\vert n\right\vert =2$.

In a conventional 2DEG, the dielectric function is given by

\begin{equation}
\varepsilon \left( \mathbf{q}\right) =1+2\left( \frac{e^{2}/\kappa \ell }{%
\hslash \omega _{c}}\right) \frac{1}{q\ell }\sum_{s}\sum_{n,n^{\prime
}}\left\vert F_{n,n^{\prime }}\left( \mathbf{q}\right) \right\vert ^{2}\frac{%
\nu _{n,s}-\nu _{n^{\prime },s}}{n^{\prime }-n}
\end{equation}%
where $n,n^{\prime }=0,1,2,...$ Since $\left( e^{2}/\kappa \ell \right)
/\hslash \omega _{c}\sim 1/\sqrt{B},$ screening is less important at large
magnetic fields in a conventional 2DEG than in graphene.

We have recomputed the energy of a Hartree-Fock electron-hole pair, $\Delta
_{e-h}^{\left( S\right) },$ a S-aS pair, $\Delta _{S-aS}^{\left( S\right) },$
as well as the $NL\sigma $ model result $\Delta _{NL\sigma M}^{\left(
S\right) }$ using the screened matrix elements. We use the superscript $(S)$
indicates the screened gaps. The gaps $\Delta _{e-h}^{\left( S\right) }$ and 
$\Delta _{NL\sigma M}^{\left( S\right) }$ are given by%
\begin{equation}
\Delta _{e-h}^{\left( S\right) }=g\mu _{B}B+\left( \frac{e^{2}}{\kappa \ell }%
\right) \int_{0}^{\infty }\frac{dx}{2\pi }\frac{\left\vert \Lambda
_{n}\left( x\right) \right\vert ^{2}}{\varepsilon \left( x\right) },
\label{bambi}
\end{equation}%
(with $\Lambda _{n}\left( x\right) $ defined in Eq. (\ref{lamn})) and%
\begin{equation}
\Delta _{NL\sigma M}^{\left( S\right) }=8\pi \rho _{s}^{\left( S\right) },
\end{equation}%
with $\int dxx^{2}\left( \ldots \right) $ replaced by $\int dx\frac{x^{2}}{%
\varepsilon \left( x\right) }\left( \ldots \right) $ in the definition of
the spin stiffness in Eqs. (\ref{ros1}) and (\ref{ros2}). We remark that, as
in the unscreened case, the gap $\Delta _{e-h}^{\left( S\right) }$ does not
depend on the angular momentum of the added electron or hole.

\begin{figure}[tbph]
\includegraphics[scale=1.0]{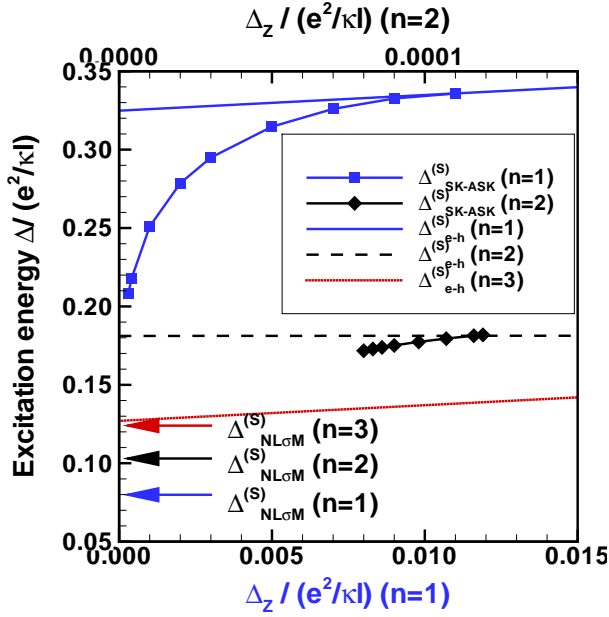}
\caption{(Color online) Excitation energy of a Hartree-Fock electron-hole
pair $\Delta _{e-h}^{\left( S\right) } $ and a skyrmion-antiskyrmion pair $%
\Delta _{S-aS}^{\left( S\right) }$ with screening corrections for Landau
levels $n=1$ (lower $x$ axis) and $n=2$ (upper $x$ axis) and $n=3$ (lower $x$
axis) at half-filling. The arrow points to the excitation energy of
skyrmion-antiskyrmion pair obtained by $NL\protect\sigma $ model with a
screened stiffness.}
\label{fig7}
\end{figure}

Fig. \ref{fig7} shows the energy gaps when screening is taken into account
for Landau levels $n=1,2,3$ in the half-filled case. The energy of all three
gaps is reduced substantially in comparison with the unscreened results. The
value of $\widetilde{g}_{c}$ is also further reduced with respect to its
unscreened value. The data points for $n=2$ are not very reliable as they
are obtained at very small Zeeman coupling where skyrmions are large. The
however provide an upper limit for $\widetilde{g}_{c}.$ The transport gap is
due to skyrmions for $n=2$ in the screened case but only at very small
Zeeman coupling. For Landau level $n=3,$ the Zeeman range of coupling where
the transport gap is due to skyrmions is further reduced with respect to the 
$n=2$ case.

\begin{figure}[tbph]
\includegraphics[scale=1.0]{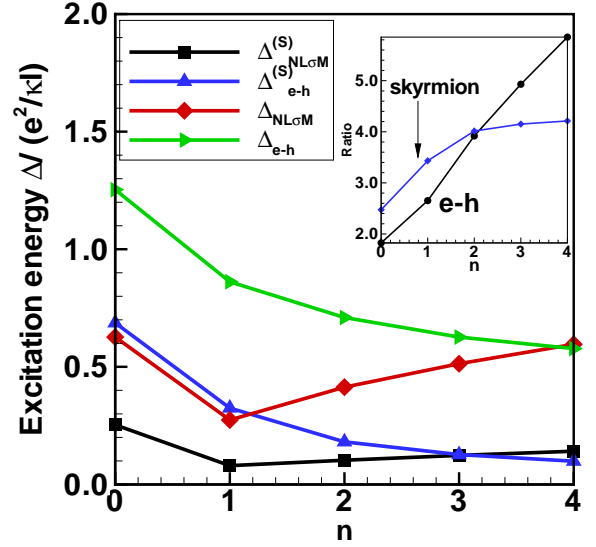}
\caption{(Color online) Evolution of the $NL\protect\sigma M$ and
electron-hole transport gaps at zero Zeeman coupling with Landau levels $%
n=1,2,3.$ The full lines are only a guide to the eyes. The inset shows the
ratios $\Delta _{e-h}/$ $\Delta _{e-h}^{(S)}$ and $\Delta _{NL\protect\sigma %
M}/$ $\Delta _{NL\protect\sigma M}^{(S)}$ with Landau level index. }
\label{fig8}
\end{figure}

Fig. \ref{fig8} shows the evolution of the $NL\sigma M$ and electron-hole
transport gaps with Landau level index at half-filling and zero Zeeman
coupling. The behavior of the ratios $\Delta _{e-h}/$ $\Delta _{e-h}^{(S)}$
and $\Delta _{NL\sigma M}/$ $\Delta _{NL\sigma M}^{(S)}$ with Landau level
index is shown in the inset. The screening corrections saturate at large $n$
more rapidly for skyrmions than for electron-hole pairs. As with $\Delta
_{NL\sigma M},$ the gap $\Delta _{NL\sigma M}^{\left( S\right) }$ increases
with Landau level index but much less rapidly than in the unscreened case.

\begin{figure}[tbph]
\includegraphics[scale=1.0]{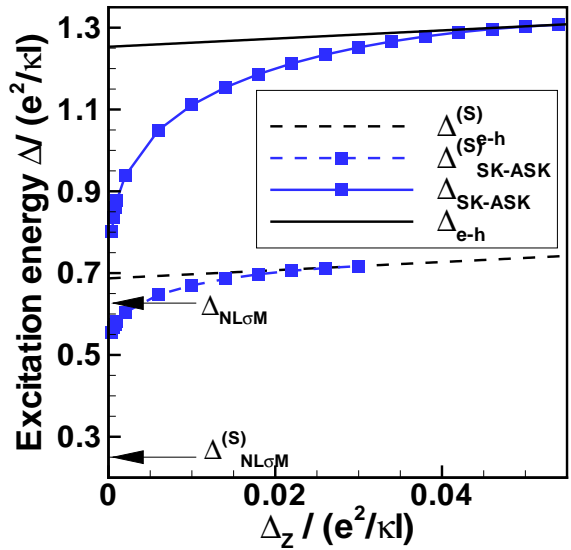}
\caption{(Color online) Excitation energy of a Hartree-Fock electron-hole
pair and a spin skyrmion-antiskyrmion pair with and without screening
corrections for Landau level $n=0$. }
\label{fig9}
\end{figure}

\begin{figure}[tbph]
\includegraphics[scale=1.0]{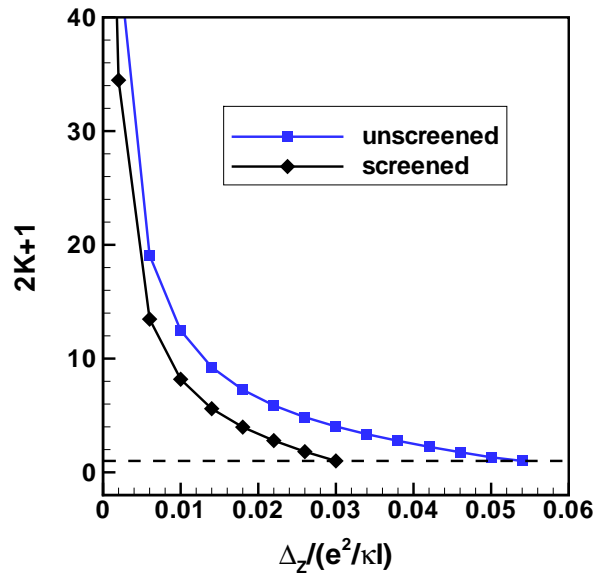}
\caption{(Color online) Number of reversed spins $N_{\downarrow }=2K+1$ in a
spin skyrmion-antiskyrmion pair in Landau level $n=0$ as a function of the
Zeeman coupling for the screened and unscreened Coulomb interaction. The
dashed line indicates $K=0.$ }
\label{fig10}
\end{figure}

For completeness, the calculation of the pair energy in both the screened
and unscreened cases for Landau level $n=0$ is shown in Fig. \ref{fig9} and
the corresponding number of down spins is shown in Fig. \ref{fig10}. As we
remarked above, the exact nature of the ground state for $n=0$ in graphene
is still controversial and is probably not spin polarized. Nevertheless, our
calculation for $n=0$ is valid for a S-aS pair energy in a conventional 2DEG.

\begin{figure}[tbph]
\includegraphics[scale=1.0]{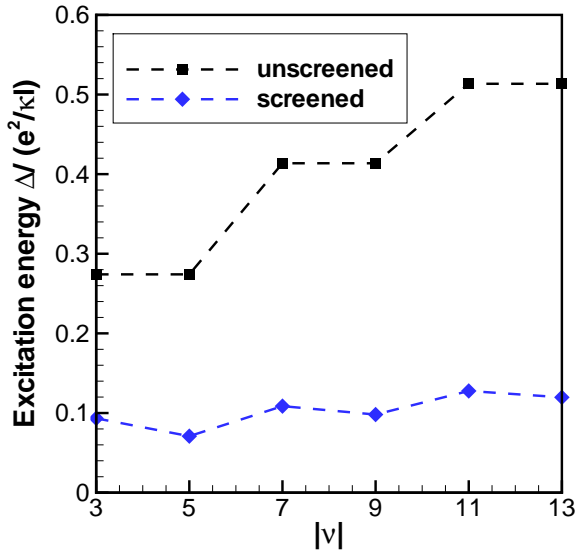}
\caption{(Color online) Excitation energy of a valley skyrmion-antiskyrmion
pair for different filling factors with and without screening corrections.
The dashed lines are only a guide to the eyes. }
\label{fig11}
\end{figure}

At $1/4$ and $3/4$ fillings of the Landau levels $\left\vert n\right\vert
\geq 1$, the ground state is valley and spin polarized. Both polarizations
are not maximal, however. At sufficiently large Zeeman coupling, spin flips
are prohibited and the lowest-energy charged excitations must be valley
skyrmions\cite{Kunyang} with up spins at $1/4$ filling and down spins at $%
3/4 $. Because there is no symmetry breaking term for the valley pseudospin,
the $NL\sigma $ model can be used to compute the S-aS excitation energy. In
the absence of screening, the gap $\Delta _{NL\sigma M}$ for valley skyrmion
at $1/4$ and $3/4$ fillings is identical to that at half-filling shown in
Fig. \ref{fig8}. As Fig. \ref{fig6} indicates, however, screening is more
important at $3/4$ filling than at $1/4$ so that we expect the transport gap
to be smaller in the former case. Fig. \ref{fig11} shows that this is indeed
the case in all Landau levels $n.$ This conclusion agrees with the
experimental results. Although there is a large sample variability in the
magnitude of the transport gap due to the different disorders, the gaps
measured at $\nu =-5$ are systematically smaller than those measured at $\nu
=-3$ (see Fig. 4 of Ref. \onlinecite{Afyoung}) by a factor $\approx 1.3$
which is close to that measured experimentally. The measurements show only a
minimal dependence of the gaps with the perpendicular magnetic field so that
our assumption of no spin flip is justified.

\section{CONCLUSION}

We have computed numerically the electron-hole and spin
skyrmion-antiskyrmion transport gaps in Landau levels $n=1$ to $n=3$ in
graphene as a function of the Zeeman coupling strength. Our calculation used
a microscopic wave function for the spin-texture excitations and the energy
was computed in the Hartree-Fock approximation. By keeping a large number of
orbital momenta (up to $m_{\max }=1000$) in the calculation, we were able to
obtain the transport gap at very small value of the Zeeman coupling $%
\widetilde{g}\approx 0.001.$

Previous calculations at zero Zeeman coupling using the nonlinear $\sigma $
model\cite{Kunyang} already indicated that the transport gap is due to spin
texture excitations in graphene at half-filling of the Landau levels $%
n=1,2,3 $ (Landau level $n=0$ is not spin polarized at half-filling) and to
valley skyrmions at $1/4$ and $3/4$ fillings. By comparison, skyrmions are
the lowest-energy charged excitations in conventional 2DEG only in Landau
level $n=0.$ Our calculations confirm the $NL\sigma M$ results and indicate
that the spin texture excitations persist for $n=1$ up to $\widetilde{g}%
_{c}\approx 0.05$ or $\widetilde{g}_{c}\approx 0.011$ when screening
corrections are included and up to $\widetilde{g}_{c}\approx 0.0026$ for $%
n=2 $ in the absence of screening. In the screened case for $n=2$ and in
both cases for $n=3,$ critical value of $\widetilde{g}_{c}$ is very small
and a reliable numerical result is difficult to obtain. Skyrmions are the
lowest-energy excitations in theses cases only in a very small range of
Zeeman coupling.

For valley skyrmions, there is no symmetry-breaking term equivalent to the
Zeeman coupling so that the transport gap can be computed using the $%
NL\sigma M$ if spin flips are prohibited by a finite Zeeman coupling. Our
results show that screening corrections are more important at $3/4$ filling
than at $1/4$ so that the transport gap due to unbound valley
skyrmion-antiskyrmion excitations is smaller in the former case.

Although screening corrections reduce substantially the size of the
transport gap, the theoretical value is still large in comparison with the
experimental result. Including disorder and Landau level mixing might help
to decrease the gap to a more realistic value. We may also consider working
with the full four-level model i.e. consider skyrmions with intertwined spin
and valley textures.

\begin{acknowledgments}
R. C\^{o}t\'{e} was supported by a grant from the Natural Sciences and
Engineering Research Council of Canada (NSERC). Computer time was provided
by Calcul Qu\'{e}bec and Compute Canada. W. Luo would like to thank Dr.
Huizhong Lu of Calcul Qu\'{e}bec for helpful discussions.
\end{acknowledgments}

\end{document}